\documentclass[aps,prl,twocolumn,amsmath,10pt,superscriptaddress,floatfix,nofootinbib]{revtex4-1}

\usepackage{amsmath}
\usepackage{epsfig}
\usepackage{color,soul}
\usepackage{multirow}

\def\bea#1\eea{\begin{align}#1\end{align}}

\newcommand{\bef}{\begin{figure}[h!tb]\centering}
\newcommand{\eef}{\end{figure}}

\begin{document}
\title{Hyperon Polarization from the Vortical Fluid in Low Energy Nuclear Collisions}

 \author{Yu Guo}
\affiliation{Guangdong Provincial Key Laboratory of Nuclear Science, Institute of Quantum Matter, 
South China Normal University, Guangzhou 510006, China.} 
\affiliation{Guangdong-Hong Kong Joint Laboratory of Quantum Matter, South China Normal University, Guangzhou 510006, China.}

 \author{Jinfeng Liao}
\email{liaoji@indiana.edu}
\affiliation{ Physics Department and Center for Exploration of Energy and Matter,
Indiana University, 2401 N Milo B. Sampson Lane, Bloomington, IN 47408, USA.}

  \author{Enke Wang} 
  \email{wangek@scnu.edu.cn}
\affiliation{Guangdong Provincial Key Laboratory of Nuclear Science, Institute of Quantum Matter, 
South China Normal University, Guangzhou 510006, China.} 
\affiliation{Guangdong-Hong Kong Joint Laboratory of Quantum Matter, South China Normal University, Guangzhou 510006, China.}

\author{Hongxi Xing}
  \email{hxing@m.scnu.edu.cn}
\affiliation{Guangdong Provincial Key Laboratory of Nuclear Science, Institute of Quantum Matter, 
South China Normal University, Guangzhou 510006, China.} 
\affiliation{Guangdong-Hong Kong Joint Laboratory of Quantum Matter, South China Normal University, Guangzhou 510006, China.}

\author{Hui Zhang}
\email{mr.zhanghui@m.scnu.edu.cn}
\affiliation{Guangdong Provincial Key Laboratory of Nuclear Science, Institute of Quantum Matter, 
South China Normal University, Guangzhou 510006, China.} 
\affiliation{Guangdong-Hong Kong Joint Laboratory of Quantum Matter, South China Normal University, Guangzhou 510006, China.}
                   
\date{\today}         

\begin{abstract}
In 2017, STAR Collaboration reported the measurements of hyperon global polarization in heavy ion collisions, suggesting the subatomic fireball fluid created in these collisions as the most vortical fluid. The results show a strongly increasing trend of the signal with decreasing beam energy down to $\sqrt{s_{NN}} = 7.7~\rm GeV$. There remains an interesting question: how the signal would change toward even lower energy. In this work we perform a systematic study on the beam energy dependence of global and local hyperon polarization phenomenon, especially in the interesting $\hat{O}(1\sim 10)\ \rm GeV$ region. We find a non-monotonic trend for the global polarization, which first increases and then decreases when beam energy is lowered from $27~\rm GeV$ down to $3~\rm GeV$. The maximum polarization signal has been identified around $\sqrt{s_{NN}} = 7.7~\rm GeV$ in this model while the averaged thermal vorticity appears to reach a peak around $3\sim4~\rm GeV$. The implications of these results are discussed. 
\end{abstract}

\date{\today}

\maketitle
\section*{Introduction}

Where can one find the fastest spinning fluid? This is a question of general interest. From a fluid dynamical point of view, the most vortical  fluid droplet needs to be large enough to enable a hydrodynamic behavior while small enough to rotate fast without violating the speed of light constraint.  In 2017, the STAR collaboration reported the measurement of hyperon global polarization in heavy ion collisions~\cite{STAR:2017ckg}, suggesting the subatomic fireball fluid created in these collisions as the most vortical fluid with an average vorticity on the order of $10^{21\sim 22} \rm s^{-1}$. This finding was further confirmed by  detailed measurements in~\cite{Adam:2018ivw,Adam:2019srw,Adam:2020pti,Acharya:2019ryw}. The observed high vorticity is a consequence of the large angular momentum carried by a small-size system in  heavy ion collisions. The angular momentum turns into nontrivial fluid vorticity patterns that lead to a global spin polarization of produced hadrons, which can be experimentally measured~\cite{Liang:2004ph,Gao:2007bc,Voloshin:2004ha,Betz:2007kg,Becattini:2007sr}. 
 Various model calculations show excellent agreement with STAR data, supporting the interpretation of the measurements in terms of rotational polarization induced by fluid vorticity~
\cite{Becattini:2014yxa,Becattini:2013vja,Csernai:2013bqa,Csernai:2014ywa,Becattini:2015ska,Becattini:2016gvu,Jiang:2016woz,Shi:2017wpk,Deng:2016gyh,Pang:2016igs,Li:2017slc,Xia:2018tes,Sun:2017xhx,Wei:2018zfb,Ayala:2020soy}.    
 See recent reviews in e.g.~\cite{Becattini:2021lfq,Gao:2020lxh,Karpenko:2021wdm,Huang:2020dtn,Becattini:2020ngo,Fukushima:2018grm,Kharzeev:2015znc,Kharzeev:2020jxw,Gao:2020vbh,Liu:2020ymh}.

One notable feature of the hyperon global polarization measurement results is a strongly increasing trend from high beam energy at $\sqrt{s_{NN}}=200 ~\rm GeV$ to low beam energy at $\sqrt{s_{NN}}= 7.7 ~\rm GeV$.  It is tempting to ask whether such a trend would continue further into the $\hat{O}(1) ~\rm GeV$ range  and   at which beam energy the truly most vortical fluid will be located. To answer these questions,  quantifying the beam energy dependence of hyperon polarization phenomenon is needed, especially in the most interesting region of $\hat{O}(1\sim 10)\ \rm GeV$ beam energy. 
{We note a number of recent studies on this problem in the literature with a variety of approaches: see~\cite{Teryaev:2015gxa, Xie:2016fjj,Baznat:2017jfj,Ivanov:2017dff,Kolomeitsev:2018svb,Ivanov:2019ern,Deng:2020ygd, Ivanov:2020udj}.}   In the present work, we carry out such a systematic beam energy scan of hyperon polarization within a transport model approach. As we will show in the rest of this paper, a non-monotonic trend for the global polarization from $27~\rm GeV$ down to $3~\rm GeV$ is found. The maximum polarization signal is located around $\sqrt{s_{NN}} = 7.7~\rm GeV$ in  this model while the averaged thermal vorticity appears to reach a peak around $3\sim4~\rm GeV$. The implications of these results will be discussed.

\section*{Formalism}

Here we present the formalism for computing the $\Lambda$ hyperon spin polarization in this work. For the overall bulk matter created in the collisions, we use the transport model AMPT~\cite{Lin:2004en,Lin:2014tya}. This model is known to provide a good description of the bulk collective dynamics such as soft particles' yields, transverse momentum spectra and flow observables, for a wide beam energy span. Efforts were also made recently for improving the model at very low beam energy to account for e.g. finite nuclear thickness~\cite{Lin:2017lcj,Mendenhall:2020fil}.  The AMPT model allows explicit tracking of every parton or hadron's motion during the evolution and of each final state hadron's formation. This enables a relatively straightforward procedure to extract the system's vorticity structure as well as to incorporate the spin polarization effect upon the hadron formation. The AMPT model  was first adapted to compute vorticity structures in  \cite{Jiang:2016woz} and  later widely used for polarization studies~\cite{Shi:2017wpk,Li:2017slc,Xia:2018tes,Guo:2019joy,Wei:2018zfb}. 
We use the same setup  as in \cite{Lin:2014tya,Jiang:2016woz} with key parameters adjusted for the low energy Au+Au collisions at RHIC. From AMPT simulations one obtains the four velocity distribution $u^\mu(x)$ as well as energy density distribution $\epsilon(x)$ over a ``mesh'' of space-time $x=(t,\vec x)$ after averaging a sufficient number of events, which can then be further used to evaluate various quantities of interest. 
We note in passing that there are other approaches for computing the vorticity and polarization in heavy ion collisions~\cite{Betz:2007kg, Bhadury:2020cop, Fukushima:2020ucl, Fu:2020oxj, Hattori:2020gqh, Gao:2020pfu, Xia:2018tes, Wei:2018zfb, Deng:2016gyh, Deng:2020ygd, Ivanov:2020udj, Teryaev:2015gxa, Xie:2016fjj, Baznat:2017jfj, Ivanov:2017dff, Kolomeitsev:2018svb, Ivanov:2019ern, Kapusta:2019ktm,Csernai:2018yok,Shi:2020htn, She:2021lhe,Li:2020eon}, such as hydrodynamic models, HIJING, URQMD, and chiral kinetic transport. 
For reviews see e.g. \cite{Huang:2020dtn, Liu:2020ymh, Becattini:2020ngo}.

The rotational polarization effect on particle spin in a relativistic fluid can be determined from the   thermal vorticity $\varpi_{\mu\nu}$, which is defined as~\cite{Becattini:2014yxa}:  
\begin{eqnarray} \label{eq_varpi}
 \varpi_{\mu \nu} = - \frac{1}{2} \left (\partial_\mu \beta_\nu - \partial_\nu \beta_\mu \right )   
\end{eqnarray}   
where $\beta_\mu = u_\mu/T$ with $T = 1/\beta$ the local temperature. A  related quantity is the kinetic vorticity defined by 
$ \Omega_{\mu\nu}= - \frac{1}{2}(\partial_\mu u_\nu - \partial_\nu u_\mu) $.
 Obviously 
 $\varpi_{\mu \nu} = \beta \left\{  \Omega_{\mu\nu} -  \left[  (\beta \partial_\mu T) u_\nu -     (\beta \partial_\nu T) u_\mu  \right]  \right\}$.  The thermal vorticity differs from the $\Omega_{\mu\nu}/T$   by terms containing gradients of temperature,  
$\sim (\beta \partial_\mu T)= [(\partial_\mu T)/T]$. While straightforward to evaluate in hydrodynamic models, such terms are trickier to compute in transport models~\cite{Shi:2017wpk,Li:2017slc,Xia:2018tes}. As a proxy, we use the energy density $\epsilon$ to evaluate such terms via $(\partial_\mu T)/T=(\partial_\mu \epsilon)/(4\epsilon)$.  

For the calculation of the hyperon polarization,  one can use the following  ensemble-averaged spin 4-vector of the produced $\Lambda$ which is  determined from the local thermal vorticity at its formation point  as~\cite{Becattini:2014yxa,STAR:2017ckg,Becattini:2016gvu,Li:2017slc,Wei:2018zfb}:  
\begin{eqnarray} \label{eq_neutral}
S^\mu =  - \frac{1}{8m}\epsilon^{\mu\nu\rho\sigma}p_\nu \varpi_{\rho\sigma} , 
\end{eqnarray} 
where $p^\nu$ and $m$ are the four-momentum and mass of each produced hyperon, respectively. 
 {
A number of remarks are in order here. Firstly, the above formula is essentially a thermal model formula for spin degree of freedom based on local thermal equilibrium. Such a working hypothesis, while well supported by experimental data and modeling analysis at relatively high energy collisions, may require more investigations to be fully vetted in the low energy regime. 
Secondly, the Eq.~(\ref{eq_neutral}) is an approximate form of a more general formula (see e.g.~\cite{Becattini:2014yxa,Becattini:2016gvu}) when the Boltzmann factor $e^{[-(E-\mu)/T]}$ is very small for the hyperons. We've numerically verified this to be true for all collision energies.   
}

In the present work, we extend the transport model calculations to the regime of very low beam energy, focusing on a systematic beam energy scan across  $\sqrt{s_{NN}} = (3\sim 10)~\rm GeV$ and examining the trend of global and local polarization observables. 
 {Specifically at each beam energy, we run AMPT simulations to collect the parton motion information and hadron kinematic freeze-out information from each event. We then use the event average to map out a profile of energy density and fluid velocity distributions, to be used for computing the thermal vorticity in Eq.~(\ref{eq_varpi}) at various spacetime points. Such information is then further used to compute the spin polarization of each hyperon via Eq.~(\ref{eq_neutral}). Finally the global and local polarization signals are computed as the  average of $S^\mu$ over all produced hyperons, in the same way as  previous studies~\cite{Li:2017slc,Shi:2017wpk,Xia:2018tes,Guo:2019joy}.} 
Our analysis here helps provide directly relevant insights for the anticipated STAR  measurement results from  the  RHIC beam energy scan program (BES-II). To ensure adequate statistics, we have simulated $10^7$  AMPT events at each beam energy we study in this work.

\begin{figure}[!hbt]
	\begin{center}
		\includegraphics[width=3.3in]{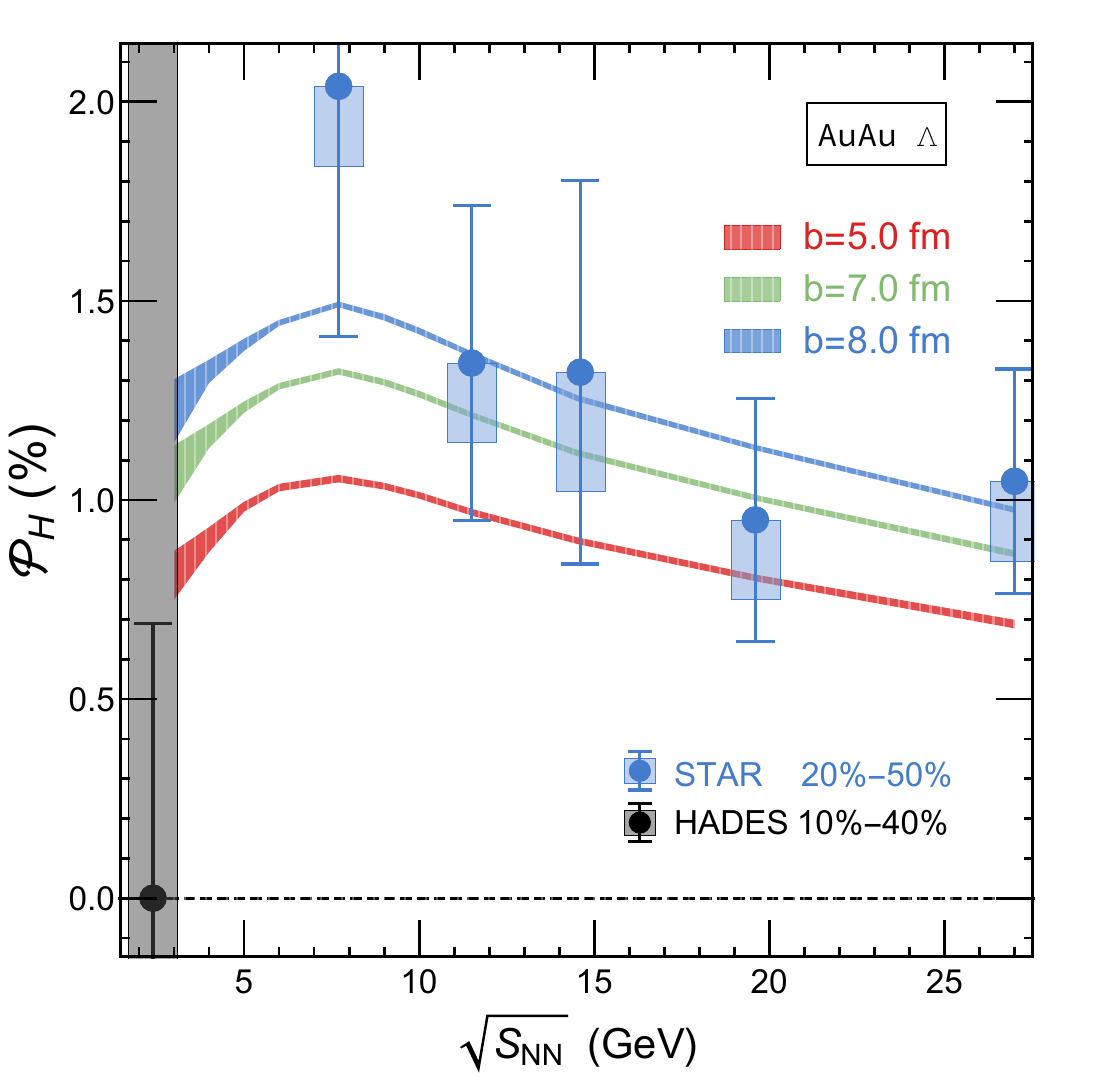}
		\caption{$\Lambda$ global polarization signal as a function of collision beam energy {with $\vert y_\Lambda\vert<1.0$ and $0.4~\rm GeV < p_T <3.0~\rm GeV$}, for AuAu collisions at three different impact parameters $b=5$ fm (red), $7$ fm (green) and $8$ fm (blue), with the bands representing statistical uncertainty from our simulation. Relevant experimental measurements are also shown, with the STAR data from Ref.\cite{STAR:2017ckg} and the HADES data from Ref. \cite{Kornas:2020qzi}. } 
		\vspace{-0.5cm}
		\label{fig_1}
	\end{center}
\end{figure}

\section*{Results}
 %
\begin{figure}[!hbt]

		\includegraphics[width=1.68in]{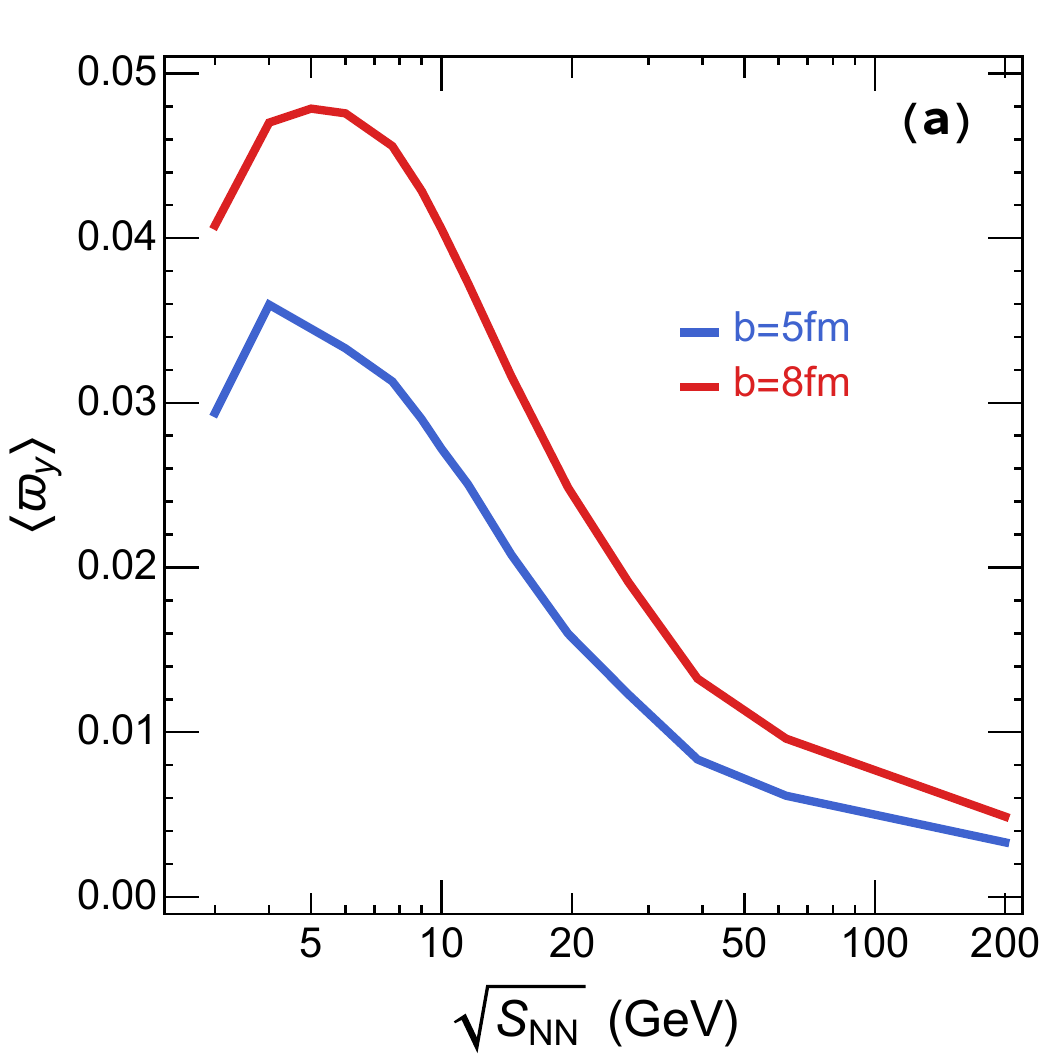}  
		\includegraphics[width=1.68in]{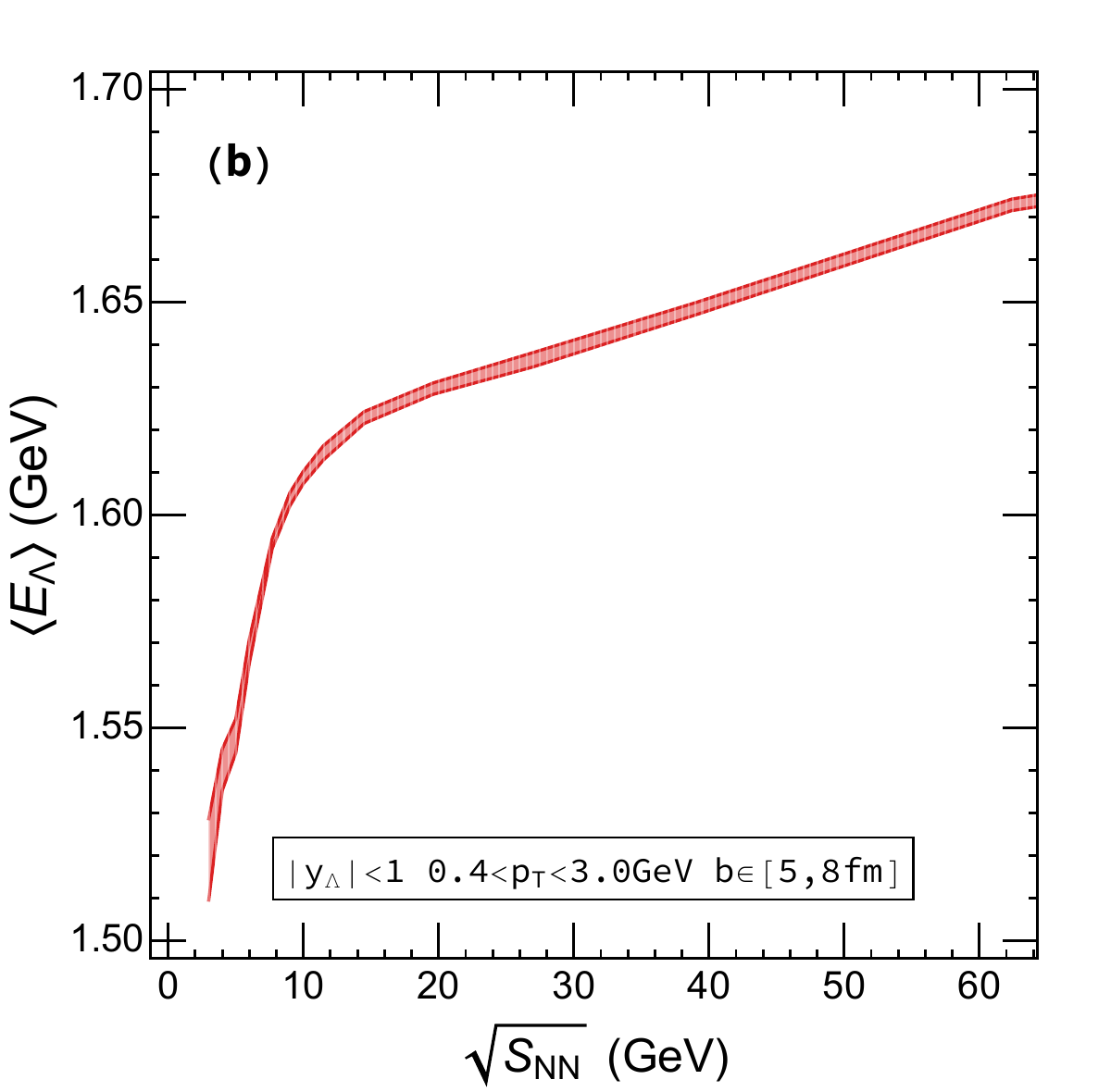}  
		\caption{{(a) The averaged thermal vorticity of the mid-rapidity ($\vert \eta\vert\le 1.0$) fireball at the fully overlapping time of two colliding nuclei as a function of beam energy for AuAu collisions with two different impact parameters $b=5$ fm (red) and $8$ fm (blue).   (b) The average energy of produced $\Lambda$ particle as a function of beam energy within kinematic window $\vert y_\Lambda\vert \le 1.0$,  $0.4~\rm GeV < p_T <3.0~\rm GeV$ and impact parameter $5~\rm fm < b <8~\rm fm$. }}
		\vspace{-0.5cm}
		\label{fig_2}
	
\end{figure}

The key finding of this work is shown in Fig.~\ref{fig_1}, where the hyperon global polarization signal is computed as a function of collision beam energy, for AuAu collisions at three different impact parameters $b=5$ fm (red), $7$ fm (green) and $8$ fm (blue). Relevant experimental measurements, where available, are also shown for comparison. We have particularly focused on the region of $\sqrt{s_{NN}} = (3\sim 10)~\rm GeV$ which has just begun to be experimentally explored.  The plot includes our  calculations  done for $\sqrt{s_{NN}} =3, 4, 5, 6, 7.7, 9, 10, 11.5, 14.5, 19.6, 27 ~\rm GeV$. This detailed and systematic scan in beam energy has allowed us to clearly identify the trend of the polarization signal. In particular, a non-monotonic is identified, with the global polarization to first increase and then decrease when beam energy is lowered from $27~\rm GeV$ down to $3~\rm GeV$. 
In the present model calculations, a maximum polarization signal is located around $\sqrt{s_{NN}} = 7.7~\rm GeV$ below which the global polarization starts to decrease. 
 {Such a decrease toward $3~\rm GeV$ is due to two important factors in the spin polarization Eq.~(\ref{eq_neutral}): (1) the thermal vorticity itself which first strongly increases and then starts to decrease past about $4\, \rm GeV$; (2) the average energy of produced hyperon particle which quickly decreases below $10\, \rm GeV$. We've explicitly verified the beam energy dependence of both factors from our simulations, with the results shown in Fig.~\ref{fig_2}. While the averaged thermal vorticity appears to reach a peak around $4~\rm GeV$, the strong downward trend in the produced hyperon energy below about $10~\rm GeV$ drives the polarization signal to decrease already around $7~\rm GeV$.  }  
 {It is informative to compare the results in this study with other models. The URQMD study in \cite{Deng:2020ygd}, while not calculating polarization directly, shows a peak of the vorticity around $3\rm GeV$ collision energy. It may be emphasized that the vorticity results from AMPT study are more or less consistent with that from URQMD model. The calculations of 3FD model in   \cite{Ivanov:2020udj}  show a peak of polarization at about $3\rm GeV$ with a signal level much larger than the AMPT result.  This may have to do with the ``frictional interaction'' between spectators and participants implemented in that specific model which may help distribute more angular momentum toward the central-rapidity fireball. }



With the anticipated upcoming experimental data for polarization signals in the $2\sim 5 ~\rm GeV$ regime, it would be interesting to test various model results with detailed measurements. As nearly all model calculations are based on fluid-vorticity-polarization scenario established at higher energy collisions, an agreement between model results and measurements could be an important confirmation of this scenario in the new regime of $\hat{O}(1) \rm GeV$ collisions. On the other hand, a discrepancy between measured signal level and modeling results could imply possible alternative mechanism of polarization generation or breakdown of certain model assumptions. Specifically concerning the AMPT model used in this work, hyperons are generated from the partonic matter via coalescence mechanism, which is certainly dominant at high collision energy but may become less important in those extremely low energy collisions where direct hadronic processes may prevail. Indeed, there is a likely shift from partonic to hadronic dominance for the bulk fluid somewhere in the $\hat{O}(1\sim 10) \rm GeV$ beam energy region. If the hyperon production mechanism in very low energy collisions becomes substantially different from the partonic picture in the AMPT model, then the present study may underestimate the polarization signal. In this sense, quantitative scan of  hyperon polarization across low beam energy  could offer valuable insights into the particle production mechanism in such collisions.

\begin{figure}[!hbt]
	\begin{center}
		\includegraphics[width=3.3in]{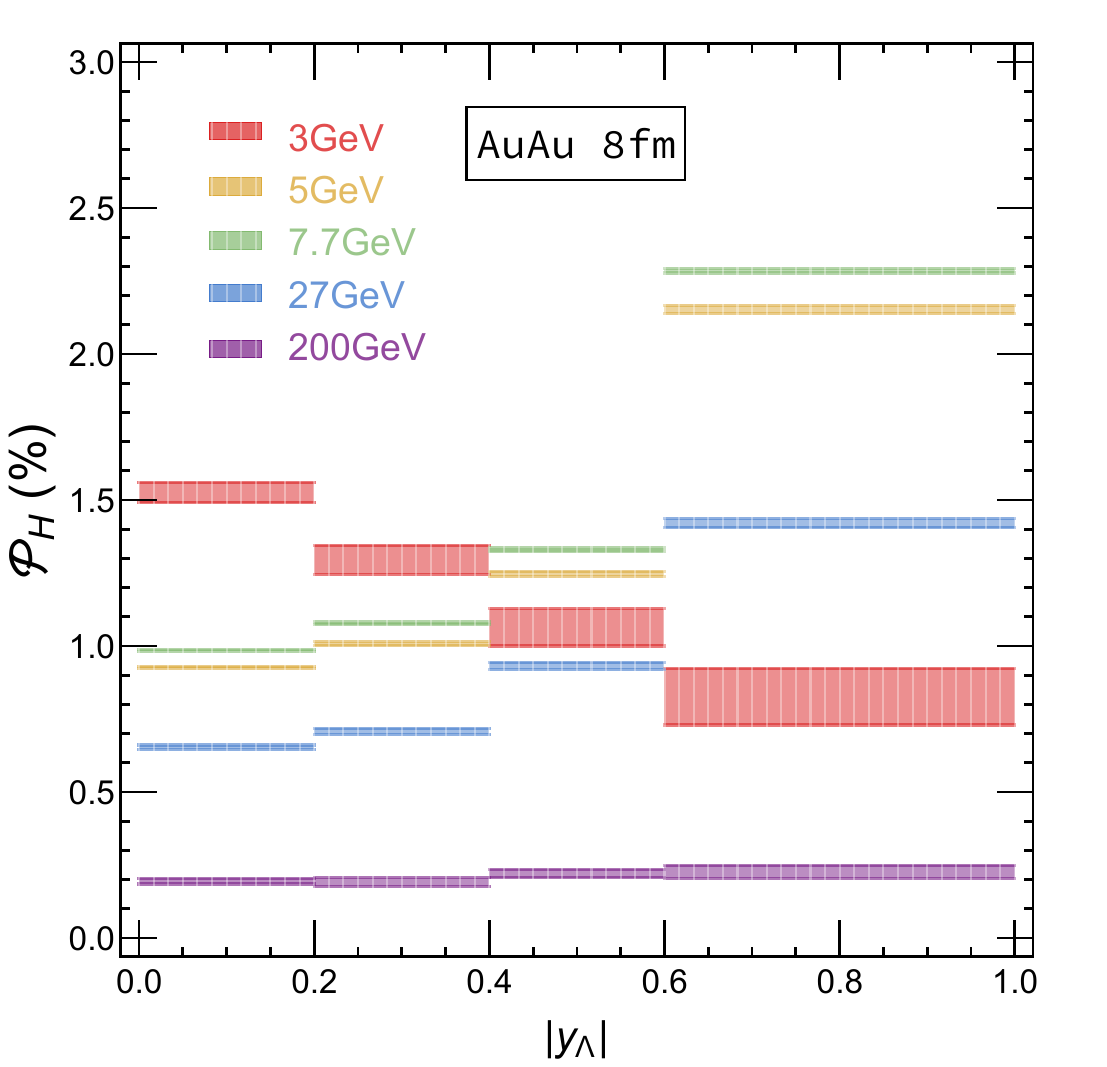}
		\caption{$\Lambda$ global polarization signal as a function of rapidity {with $0.4~\rm GeV < p_T <3.0~\rm GeV$}, for AuAu collisions at  impact parameter $b=8$ fm for a variety of collision beam energy from $200\rm GeV$ down to $3\rm GeV$, with the bands representing statistical uncertainty from our simulation. }
		\vspace{-0.5cm}
		\label{fig_3}
	\end{center}
\end{figure}

\begin{figure}[!hbt]
	\begin{center}
		\includegraphics[width=3.3in]{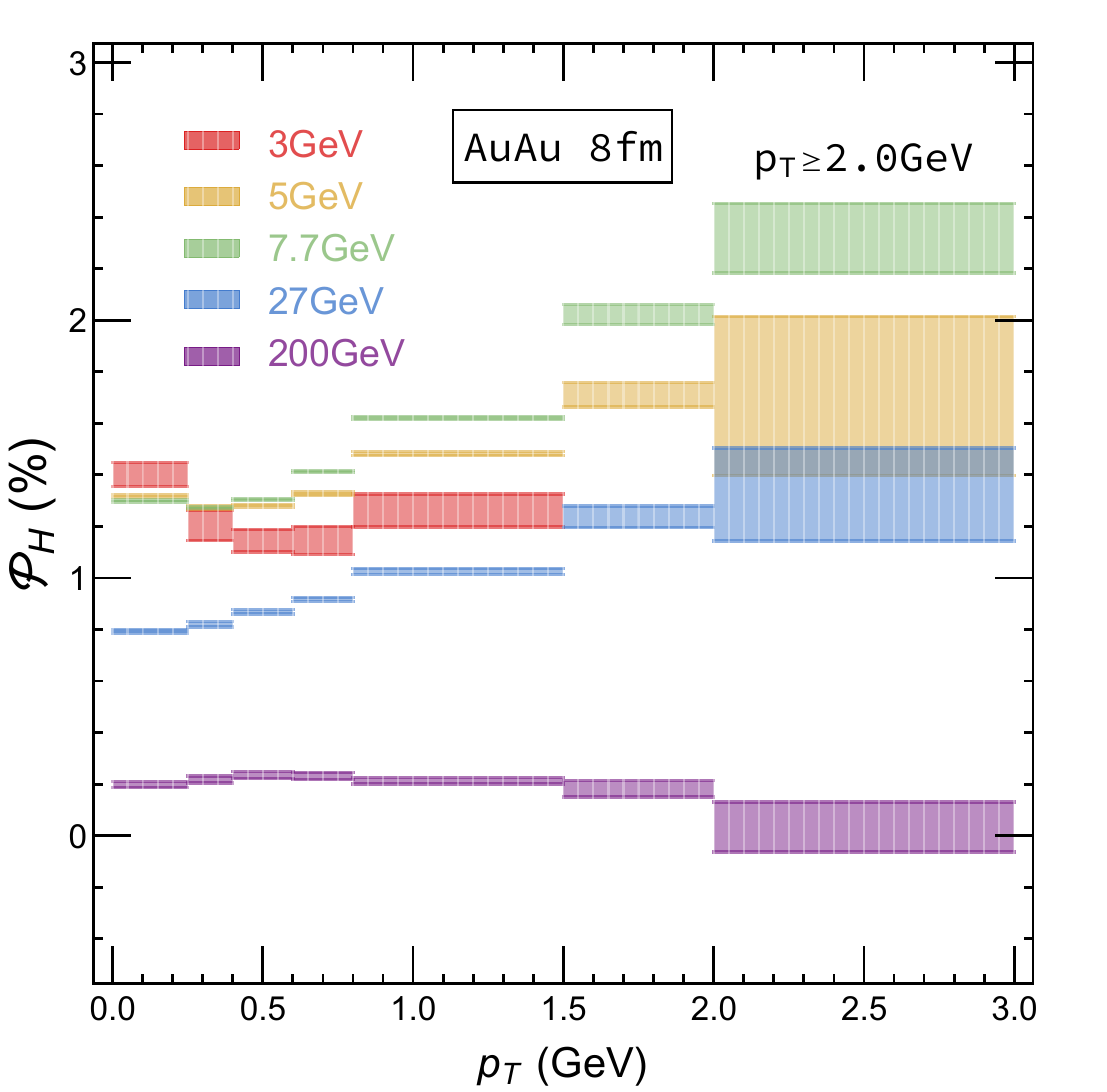}   
		\caption{$\Lambda$ global polarization signal as a function of transverse momentum {with $\vert y_\Lambda\vert<1.0$}, for AuAu collisions at  impact parameter $b=8$ fm for a variety of collision beam energy from $200 ~\rm GeV$ down to $3 ~\rm GeV$, , with the bands representing statistical uncertainty from our simulation.  }
		\vspace{-0.5cm}
		\label{fig_4}
	\end{center}
\end{figure}

We've also computed the differential observables of  hyperon global polarization over a wide beam energy span.   The dependence  on  the rapidity is shown in Fig.~\ref{fig_3}. A barely visible increase with rapidity at $200~\rm GeV$ changes into a strong rising behavior at low energy down to $5~ \rm GeV$, which then turns into a strong decrease with rapidity at $3 ~\rm GeV$. This might be understood from the dependence of vorticity on rapidity, which typically increases first and then decreases, while the shift  from high beam energy to low beam energy has an ``inward-shrinking'' effect on the rapidity dependence. The dependence on the   transverse momentum  in Fig.~\ref{fig_4} also shows a similar pattern, in which both the vorticity and the radial flow may play a role due to their different dependence on beam energy.   

\begin{figure}[!hbt]
	\begin{center}
		\includegraphics[width=3.3in]{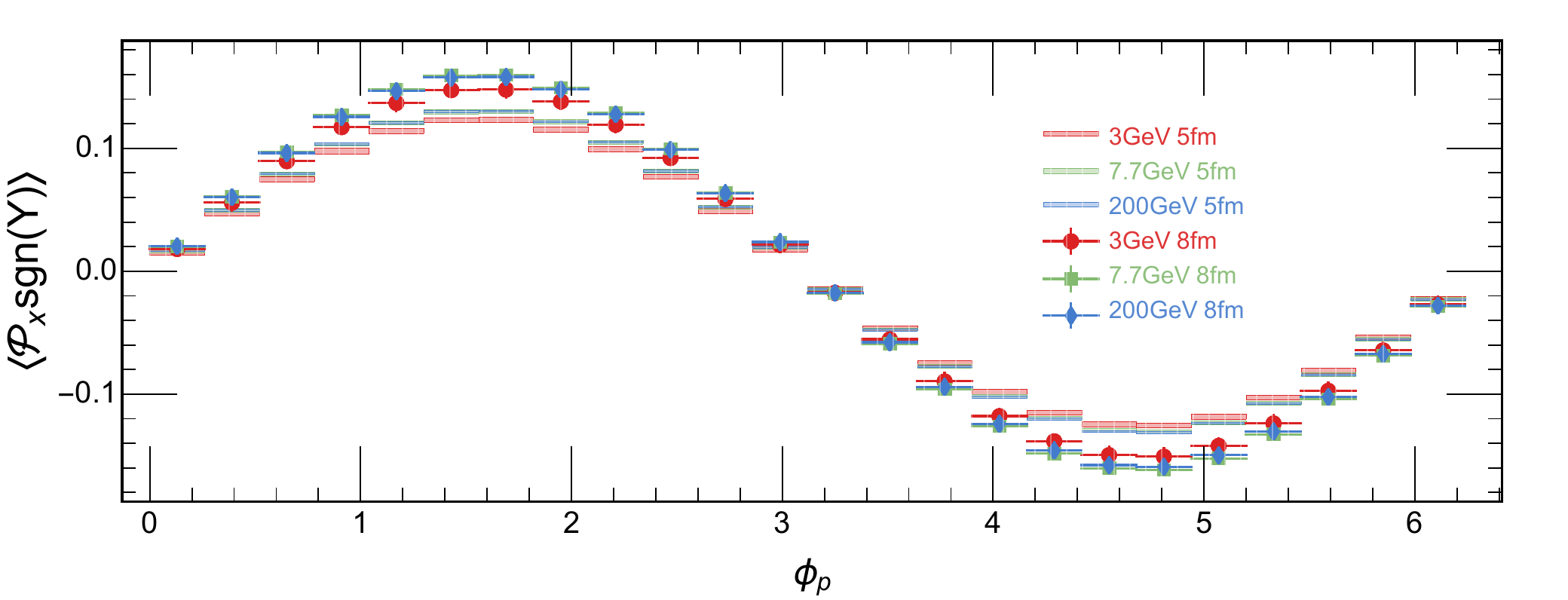}  
		\includegraphics[width=3.3in]{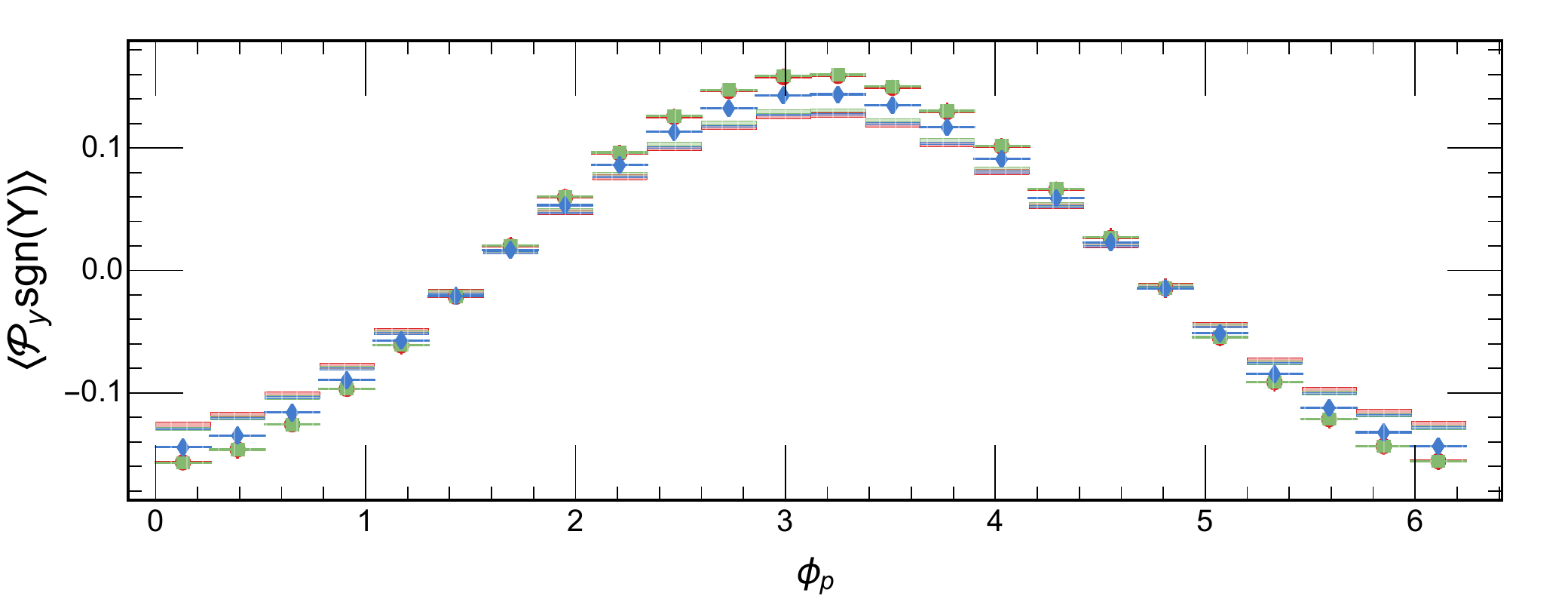}  
		\includegraphics[width=3.3in]{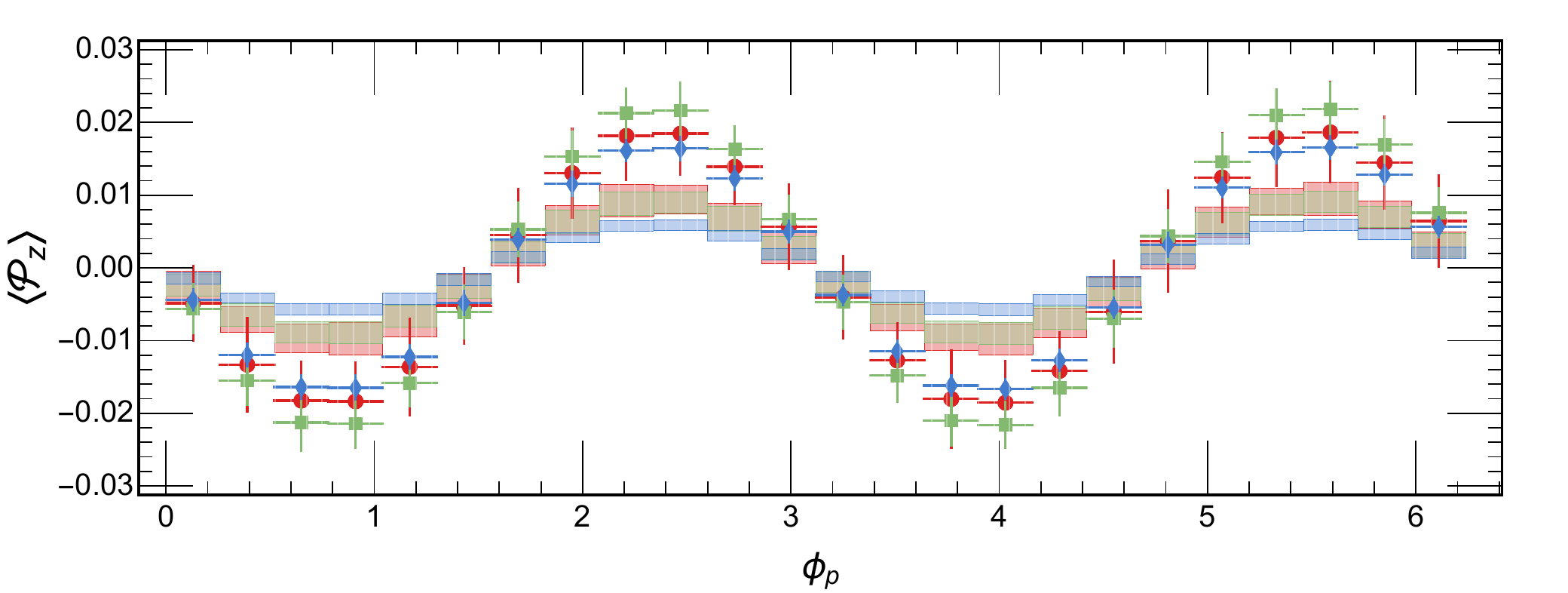}   
		\caption{The dependence of $\Lambda$ local polarization components on the azimuthal angle $\phi_p$ for collisions with impact parameter $b=5, 8 \ \rm fm$ and with  beam energy $\sqrt{s_{NN}}= 3, 7.7, 200 \ \rm GeV$ {with $\vert y_\Lambda\vert<1.0$ and $0.4~\rm GeV < p_T <3.0~\rm GeV$}, respectively.  The bands (for $b=5\rm fm$ results) or error bars (for $b=8\rm fm$ results) represent statistical uncertainty from our simulation.}
		\vspace{-0.5cm}
		\label{fig_5}
	\end{center}
\end{figure}

In addition to global polarization, the local polarization patterns of produced hyperons provide another category of interesting observables~\cite{Becattini:2017gcx}. Measurements by STAR Collaboration in $200 \rm GeV$ collisions~\cite{Adam:2019srw}  show an azimuthal pattern of the longitudinal polarization that contradicts a large class of model calculations based on either hydrodynamic or transport models. This tension points to certain potentially subtle effect beyond the global polarization in thermal equilibrium. It has recently been argued that~\cite{Becattini:2021iol,Becattini:2021suc,Fu:2021pok,Liu:2021uhn}  a nontrivial coupling between spin and thermal shear might help resolve the puzzle observed in $200 \rm GeV$ collisions, which might be absent in low energy collisions. Clearly, it is of great interest to calculate the local polarization patterns for low energy collisions.  In Fig.~\ref{fig_5}, we show the dependence of all three components of the hyperon local polarization on the azimuthal angle $\phi_p$ with impact parameter $b=5, 8 \ \rm fm$, which are the first results of its kind for collision beam energies at $3$ and $7.7$ $\rm GeV$. The results suggest that these azimuthal angular patterns are relatively insensitive to the collisional beam energy, possibly because the local vorticity patterns are dominated by the radial flow of the fireball and less sensitive to the overall angular momentum~\cite{Jiang:2016woz}.  These results can be compared with future measurements at low energy to offer insights on the data-model discrepancy seen at high beam energy.

\section*{Conclusion}

In summary, we've performed a systematic study on the beam energy dependence of hyperon global polarization phenomenon in heavy ion collisions, especially focusing on the interesting $\hat{O}(1\sim 10)\ \rm GeV$ region. Our key finding is a non-monotonic trend, with the global polarization to first increase and then decrease when beam energy is lowered from $27~\rm GeV$ down to $3~\rm GeV$. The maximum polarization signal has been identified around $\sqrt{s_{NN}} = 7.7~\rm GeV$, where the heavy ion collisions presumably create the most vortical fluid. We've also reported differential dependence of hyperon global polarization on rapidity and on transverse momentum over a wide beam energy span, suggesting an interesting change of patterns  between collisions at $\hat{O}(10)~\rm GeV$, and at $\hat{O}(1)~\rm GeV$. On the other hand, our calculations for the azimuthal patterns of hyperon local polarization components demonstrate very little change across a wide span of beam energies.  Detailed experimental measurements in the $\hat{O}(1\sim 10)\ \rm GeV$ beam energy region are expected to become available very soon, which shall provide interesting test of the results from the present work as well as from other relevant model studies and offer unique insights into the   vorticity and polarization phenomenon in the low energy nuclear collisions.

\vspace{-0.1in}
\section*{Acknowledgments}
We thank J. Adams, Z.-W. Lin M. Lisa, S. Shi and N. Xu for helpful discussions and communications. This work is supported in part by the Guangdong Major Project of Basic and Applied Basic Research (No. 2020B0301030008), by the Key Project of Science and Technology of Guangzhou (Grant No. 2019050001), by the National Natural Science Foundation of China under Grants No. 12022512, No. 12035007, No. 12047523,  as well as by the NSF Grant No. PHY-1913729 and the U.S. Department of Energy, Office of Science, Office of Nuclear Physics, within the framework of the Beam Energy Scan Theory (BEST) Topical Collaboration.

\bibliographystyle{h-physrev5}   
\bibliography{biblio}

\begin{thebibliography}{10}

\bibitem{STAR:2017ckg}
STAR, L.~Adamczyk {\em et~al.},
\newblock Nature {\bf 548}, 62 (2017), arXiv:1701.06657.

\bibitem{Adam:2018ivw}
STAR, J.~Adam {\em et~al.},
\newblock Phys. Rev. C {\bf 98}, 014910 (2018), arXiv:1805.04400.

\bibitem{Adam:2019srw}
STAR, J.~Adam {\em et~al.},
\newblock Phys. Rev. Lett. {\bf 123}, 132301 (2019), arXiv:1905.11917.

\bibitem{Adam:2020pti}
STAR, J.~Adam {\em et~al.},
\newblock Phys. Rev. Lett. {\bf 126}, 162301 (2021), arXiv:2012.13601.

\bibitem{Acharya:2019ryw}
ALICE, S.~Acharya {\em et~al.},
\newblock Phys. Rev. C {\bf 101}, 044611 (2020), arXiv:1909.01281.

\bibitem{Liang:2004ph}
Z.-T. Liang and X.-N. Wang,
\newblock Phys. Rev. Lett. {\bf 94}, 102301 (2005), arXiv:nucl-th/0410079,
\newblock [Erratum: Phys.Rev.Lett. 96, 039901 (2006)].

\bibitem{Gao:2007bc}
J.-H. Gao {\em et~al.},
\newblock Phys. Rev. C {\bf 77}, 044902 (2008), arXiv:0710.2943.

\bibitem{Voloshin:2004ha}
S.~A. Voloshin,
\newblock (2004), arXiv:nucl-th/0410089.

\bibitem{Betz:2007kg}
B.~Betz, M.~Gyulassy, and G.~Torrieri,
\newblock Phys. Rev. C {\bf 76}, 044901 (2007), arXiv:0708.0035.

\bibitem{Becattini:2007sr}
F.~Becattini, F.~Piccinini, and J.~Rizzo,
\newblock Phys. Rev. C {\bf 77}, 024906 (2008), arXiv:0711.1253.

\bibitem{Becattini:2014yxa}
F.~Becattini, L.~Bucciantini, E.~Grossi, and L.~Tinti,
\newblock Eur. Phys. J. C {\bf 75}, 191 (2015), arXiv:1403.6265.

\bibitem{Becattini:2013vja}
F.~Becattini, L.~Csernai, and D.~J. Wang,
\newblock Phys. Rev. C {\bf 88}, 034905 (2013), arXiv:1304.4427,
\newblock [Erratum: Phys.Rev.C 93, 069901 (2016)].

\bibitem{Csernai:2013bqa}
L.~P. Csernai, V.~K. Magas, and D.~J. Wang,
\newblock Phys. Rev. C {\bf 87}, 034906 (2013), arXiv:1302.5310.

\bibitem{Csernai:2014ywa}
L.~P. Csernai, D.~J. Wang, M.~Bleicher, and H.~St\"ocker,
\newblock Phys. Rev. C {\bf 90}, 021904 (2014).

\bibitem{Becattini:2015ska}
F.~Becattini {\em et~al.},
\newblock Eur. Phys. J. C {\bf 75}, 406 (2015), arXiv:1501.04468,
\newblock [Erratum: Eur.Phys.J.C 78, 354 (2018)].

\bibitem{Becattini:2016gvu}
F.~Becattini, I.~Karpenko, M.~Lisa, I.~Upsal, and S.~Voloshin,
\newblock Phys. Rev. C {\bf 95}, 054902 (2017), arXiv:1610.02506.

\bibitem{Jiang:2016woz}
Y.~Jiang, Z.-W. Lin, and J.~Liao,
\newblock Phys. Rev. C {\bf 94}, 044910 (2016), arXiv:1602.06580,
\newblock [Erratum: Phys.Rev.C 95, 049904 (2017)].

\bibitem{Shi:2017wpk}
S.~Shi, K.~Li, and J.~Liao,
\newblock Phys. Lett. B {\bf 788}, 409 (2019), arXiv:1712.00878.

\bibitem{Deng:2016gyh}
W.-T. Deng and X.-G. Huang,
\newblock Phys. Rev. C {\bf 93}, 064907 (2016), arXiv:1603.06117.

\bibitem{Pang:2016igs}
L.-G. Pang, H.~Petersen, Q.~Wang, and X.-N. Wang,
\newblock Phys. Rev. Lett. {\bf 117}, 192301 (2016), arXiv:1605.04024.

\bibitem{Li:2017slc}
H.~Li, L.-G. Pang, Q.~Wang, and X.-L. Xia,
\newblock Phys. Rev. C {\bf 96}, 054908 (2017), arXiv:1704.01507.

\bibitem{Xia:2018tes}
X.-L. Xia, H.~Li, Z.-B. Tang, and Q.~Wang,
\newblock Phys. Rev. C {\bf 98}, 024905 (2018), arXiv:1803.00867.

\bibitem{Sun:2017xhx}
Y.~Sun and C.~M. Ko,
\newblock Phys. Rev. C {\bf 96}, 024906 (2017), arXiv:1706.09467.

\bibitem{Wei:2018zfb}
D.-X. Wei, W.-T. Deng, and X.-G. Huang,
\newblock Phys. Rev. C {\bf 99}, 014905 (2019), arXiv:1810.00151.

\bibitem{Ayala:2020soy}
A.~Ayala {\em et~al.},
\newblock Phys. Lett. B {\bf 810}, 135818 (2020), arXiv:2003.13757.

\bibitem{Becattini:2021lfq}
F.~Becattini, J.~Liao, and M.~Lisa,
\newblock (2021), arXiv:2102.00933.

\bibitem{Gao:2020lxh}
J.-H. Gao, Z.-T. Liang, Q.~Wang, and X.-N. Wang,
\newblock (2020), arXiv:2009.04803.

\bibitem{Karpenko:2021wdm}
I.~Karpenko,
\newblock {Vorticity and Polarization in Heavy Ion Collisions: Hydrodynamic
  Models},
\newblock 2021, arXiv:2101.04963.

\bibitem{Huang:2020dtn}
X.-G. Huang, J.~Liao, Q.~Wang, and X.-L. Xia,
\newblock (2020), arXiv:2010.08937.

\bibitem{Becattini:2020ngo}
F.~Becattini and M.~A. Lisa,
\newblock Ann. Rev. Nucl. Part. Sci. {\bf 70}, 395 (2020), arXiv:2003.03640.

\bibitem{Fukushima:2018grm}
K.~Fukushima,
\newblock Prog. Part. Nucl. Phys. {\bf 107}, 167 (2019), arXiv:1812.08886.

\bibitem{Kharzeev:2015znc}
D.~E. Kharzeev, J.~Liao, S.~A. Voloshin, and G.~Wang,
\newblock Prog. Part. Nucl. Phys. {\bf 88}, 1 (2016), arXiv:1511.04050.

\bibitem{Kharzeev:2020jxw}
D.~E. Kharzeev and J.~Liao,
\newblock Nature Rev. Phys. {\bf 3}, 55 (2021), arXiv:2102.06623.

\bibitem{Gao:2020vbh}
J.-H. Gao, G.-L. Ma, S.~Pu, and Q.~Wang,
\newblock Nucl. Sci. Tech. {\bf 31}, 90 (2020), arXiv:2005.10432.

\bibitem{Liu:2020ymh}
Y.-C. Liu and X.-G. Huang,
\newblock Nucl. Sci. Tech. {\bf 31}, 56 (2020), arXiv:2003.12482.

\bibitem{Teryaev:2015gxa}
O.~Teryaev and R.~Usubov,
\newblock Phys. Rev. C {\bf 92}, 014906 (2015).

\bibitem{Xie:2016fjj}
Y.~L. Xie, M.~Bleicher, H.~St\"ocker, D.~J. Wang, and L.~P. Csernai,
\newblock Phys. Rev. C {\bf 94}, 054907 (2016), arXiv:1610.08678.

\bibitem{Baznat:2017jfj}
M.~Baznat, K.~Gudima, A.~Sorin, and O.~Teryaev,
\newblock Phys. Rev. C {\bf 97}, 041902 (2018), arXiv:1701.00923.

\bibitem{Ivanov:2017dff}
Y.~B. Ivanov and A.~A. Soldatov,
\newblock Phys. Rev. C {\bf 95}, 054915 (2017), arXiv:1701.01319.

\bibitem{Kolomeitsev:2018svb}
E.~E. Kolomeitsev, V.~D. Toneev, and V.~Voronyuk,
\newblock Phys. Rev. C {\bf 97}, 064902 (2018), arXiv:1801.07610.

\bibitem{Ivanov:2019ern}
Y.~B. Ivanov, V.~D. Toneev, and A.~A. Soldatov,
\newblock Phys. Rev. C {\bf 100}, 014908 (2019), arXiv:1903.05455.

\bibitem{Deng:2020ygd}
X.-G. Deng, X.-G. Huang, Y.-G. Ma, and S.~Zhang,
\newblock Phys. Rev. C {\bf 101}, 064908 (2020), arXiv:2001.01371.

\bibitem{Ivanov:2020udj}
Y.~B. Ivanov,
\newblock Phys. Rev. C {\bf 103}, L031903 (2021), arXiv:2012.07597.

\bibitem{Lin:2004en}
Z.-W. Lin, C.~M. Ko, B.-A. Li, B.~Zhang, and S.~Pal,
\newblock Phys. Rev. C {\bf 72}, 064901 (2005), arXiv:nucl-th/0411110.

\bibitem{Lin:2014tya}
Z.-W. Lin,
\newblock Phys. Rev. C {\bf 90}, 014904 (2014), arXiv:1403.6321.

\bibitem{Lin:2017lcj}
Z.-W. Lin,
\newblock Phys. Rev. C {\bf 98}, 034908 (2018), arXiv:1704.08418.

\bibitem{Mendenhall:2020fil}
T.~Mendenhall and Z.-W. Lin,
\newblock Phys. Rev. C {\bf 103}, 024907 (2021), arXiv:2012.13825.

\bibitem{Guo:2019joy}
Y.~Guo, S.~Shi, S.~Feng, and J.~Liao,
\newblock Phys. Lett. B {\bf 798}, 134929 (2019), arXiv:1905.12613.

\bibitem{Bhadury:2020cop}
S.~Bhadury, W.~Florkowski, A.~Jaiswal, A.~Kumar, and R.~Ryblewski,
\newblock Phys. Rev. D {\bf 103}, 014030 (2021), arXiv:2008.10976.

\bibitem{Fukushima:2020ucl}
K.~Fukushima and S.~Pu,
\newblock Phys. Lett. B {\bf 817}, 136346 (2021), arXiv:2010.01608.

\bibitem{Fu:2020oxj}
B.~Fu, K.~Xu, X.-G. Huang, and H.~Song,
\newblock Phys. Rev. C {\bf 103}, 024903 (2021), arXiv:2011.03740.

\bibitem{Hattori:2020gqh}
K.~Hattori, Y.~Hidaka, N.~Yamamoto, and D.-L. Yang,
\newblock JHEP {\bf 02}, 001 (2021), arXiv:2010.13368.

\bibitem{Gao:2020pfu}
J.-H. Gao, Z.-T. Liang, and Q.~Wang,
\newblock Int. J. Mod. Phys. A {\bf 36}, 2130001 (2021), arXiv:2011.02629.

\bibitem{Kapusta:2019ktm}
J.~I. Kapusta, E.~Rrapaj, and S.~Rudaz,
\newblock Phys. Rev. C {\bf 101}, 031901 (2020), arXiv:1910.12759.

\bibitem{Csernai:2018yok}
L.~P. Csernai, J.~I. Kapusta, and T.~Welle,
\newblock Phys. Rev. C {\bf 99}, 021901 (2019), arXiv:1807.11521.

\bibitem{Shi:2020htn}
S.~Shi, C.~Gale, and S.~Jeon,
\newblock Phys. Rev. C {\bf 103}, 044906 (2021), arXiv:2008.08618.

\bibitem{She:2021lhe}
D.~She, A.~Huang, D.~Hou, and J.~Liao,
\newblock (2021), arXiv:2105.04060.

\bibitem{Li:2020eon}
S.~Li, M.~A. Stephanov, and H.-U. Yee,
\newblock Phys. Rev. Lett. {\bf 127}, 082302 (2021), arXiv:2011.12318.

\bibitem{Kornas:2020qzi}
HADES, F.~J. Kornas,
\newblock Springer Proc. Phys. {\bf 250}, 435 (2020).

\bibitem{Becattini:2017gcx}
F.~Becattini and I.~Karpenko,
\newblock Phys. Rev. Lett. {\bf 120}, 012302 (2018), arXiv:1707.07984.

\bibitem{Becattini:2021iol}
F.~Becattini, M.~Buzzegoli, A.~Palermo, G.~Inghirami, and I.~Karpenko,
\newblock (2021), arXiv:2103.14621.

\bibitem{Becattini:2021suc}
F.~Becattini, M.~Buzzegoli, and A.~Palermo,
\newblock (2021), arXiv:2103.10917.

\bibitem{Fu:2021pok}
B.~Fu, S.~Y.~F. Liu, L.~Pang, H.~Song, and Y.~Yin,
\newblock (2021), arXiv:2103.10403.

\bibitem{Liu:2021uhn}
S.~Y.~F. Liu and Y.~Yin,
\newblock (2021), arXiv:2103.09200.

\end{thebibliography}

\end{document}